\pdfoutput=1
\documentclass[11pt]{article}
\usepackage{graphicx}
\usepackage{epstopdf} 
\usepackage{helvet}
\usepackage{setspace}
\usepackage{rotating}
\usepackage{booktabs}
\usepackage{lscape}
\usepackage[superscript]{cite}
\usepackage{caption}
\usepackage{longtable}
\usepackage{color}
\begin{document}
\thispagestyle{empty}
\fontfamily{phv}\selectfont{
{\LARGE \noindent Towards a scientifically tenable description of\\ objective idealism}
\\
\\
{\scriptsize Martin Korth$^{*}$\\
IVV NWZ, WWU Münster, Wilhelm-Klemm-Str. 10, 48149 Münster (Germany)\\
$^{*}$ Corresponding author email: dgd@uni-muenster.de\\
v1, 2022/08/22}\\ \\



\noindent
The tremendous advances of research into artificial intelligence as well as neuroscience made over the last two to three decades 
have given further support to a renewed interest into philosophical discussions of the mind-body problem. 
Especially the last decade has seen a revival of panpsychist and idealist considerations, 
often focused on solving philosophical puzzles like the so-called hard problem of consciousness.
\cite{Foster1982,Sprigge1984,Hosle1986,Nagel1976,Chalmers1996,Bruntrup2008,BruntrupJaskolla2016,Goff2017,GoffMoran2021} 
While a number of respectable philosophers advocate some sort of panpsychistic solution to the mind-body problem now,
fewer advocate that idealism can contribute substantially to the debate.
Interest in idealism has nevertheless risen again, as can be seen also from recent overview articles and collections of works.
\cite{Meixner2016,Chalmers2019,Idealism2017,HosleMuller,IdealSci2020}
The working hypothesis here is that a properly formulated idealism can not only provide an alternative view of the mind/matter gap,
but that this new view will also shed light on open questions in our common scientific, i.e. materialist, world view.
To investigate this possibility, idealism first of all needs a model for the integration of modern science which allows for a mathematically consistent re-interpretation
of the physical world as a limiting case of a both material and non-material world, which would make the outcome of idealistic considerations accessible to scientific investigation.
To develop such a model I will first try to explain what I mean when I speak of a `scientifically tenable' idealism,
including a formulation of the emanation problem which for idealism replaces the interaction problem,
then give a very brief summary of the available elements of such a theory in the philosophical literature,
before sketching out some `design questions' which have to be answered upon the construction of such models,
and finally put forward a first model for a scientifically tenable objective idealism.
\\ \\

\noindent {\bf \fontfamily{phv}\selectfont{A scientifically-tenable idealism}}\\
As can be seen from the literature overview below, idealism has for some time retreated into a position
with few options for a constructive, two-sided exchange with modern science.
This way it was and is still able to serve as some kind of background stories for other investigations,
especially into ethics and aesthetics, and in a very abstract sense even science,
but the idea that idealism explains fundamentals of reality unavailable to scientific endeavors,
which are therefore restricted to subsequent ground work, is a vacuous claim if no real interplay occurs.
From the view-point of science such background stories are bound to have a religious feel to them,
and more importantly, such a view of idealism is as impotent as materialism to explain the mind body problem.
A scientifically tenable idealism would therefore be a formulation of idealism
which puts in concrete terms how we have to imagine the emergence of matter from non-matter,
thereby allowing for combined philosophical and scientific investigations,
with the goal of making predictions about rationally accessible consequences.
(Most likely, to have a chance of success, many wrong proposals will have to go down in history first -- so here we go ...)
The touchstone for whether idealism has something to contribute would then be,
whether idealism can help us to form a more consistent picture of reality,
for instance concerning the functioning of our brain, the measurement problem in quantum theory, etc.
Chalmers and McQueen have made an excellent (though hitherto unsuccessful) attempt at this in the context of panpsychism.\cite{ChalmersQM}
A direct transfer of this idea is not possible
as for the panpsychist the measurement problem might reveal a space for interaction of consciousness with the physical world,
while for the objective idealist it is just a real `measurement' of preparing a cut-out of our both material and non-material reality.
But to succeed, idealism will need similar ideas.
\\ \\

\noindent {\bf \fontfamily{phv}\selectfont{The emanation problem}}\\
The core problem for -- and accordingly most interesting bit of -- a scientifically tenable idealism has thus quite some overlap,
but is not the same as the interaction problem of dualism, nor the combination problem of panpsychism.
Referring back to Plotinus and especially Proclus we could call it `the emanation problem',
of how exactly the emergence of matter from non-matter works, including material causality, and in line with modern science.
It should be clear that for a long time in the history of idealist thought this problem was not at the center of attention,
as the overwhelmingly successful unification of the theories of modern physics, which is at the heart of the problem,
happened only within the last two centuries.
And although idealists might still claim that the bridge is there, but principally not further accessible to rational inquiry,
they would have to counter the argument that this is an effectively dualist position
that has little to offer for modern science and/or the mind/body problem.
In the following I will start my considerations from the viewpoint of objective idealism,
assuming the objective existence of non-material building blocks of reality,
as opposed to subjective idealism, i.e. that the world is solely the product of interacting singular mind(s),
because the unavailability of objectively existing entities in the later has no advantages,
but introduces additional problems for the formulation of our model.
\\ \\

\noindent {\bf \fontfamily{phv}\selectfont{Historical answers}}\\
\noindent 
Objective idealism can be understood to start with Plato's theory of forms, in which objectively existing, but non-material `ideas' allow the structuring of matter into bodies.
This structuring is done by some active element which gives form, i.e. the shapes of Platonic bodies, to matter, i.e. triangular corpuscles of the classical elements,
to generate the basic building blocks of the material world.

The thinkers of Middle Platonism kept Plato's theories alive, before Plotinus and in his succession the Neoplatonists, and especially Proclus, worked out a detailed structure to incorporate Plato's thinking,
proposing four `hypostasis' or layers of fundamental reality from the One (as overall unifying principle) to the Intellect or {\it Nous} (as the realm of ideas) and the (World-)Soul,
all three beyond the final layer of the world of bodies. 
In these theories, some active part of the world-soul is responsible for the structuring of material reality,
still pretty much in the sense of Plato's thinking.
Neoplatonism had great influence on thinkers of Jewish, Islamic, as well as Christian tradition.
Parts of Plotinus' {\it Enneads} circulated in the Arab world as {\it The Theology of Aristotele}
and parts of Proclus {\it Elements of Theology} as the {\it Book of Causes}, both falsely attributed to Aristotle,
with readers such as Al-Kindi, Al-Farabi and Avicenna.
The rich influence of Neoplatonism on the Christian tradition probably peaked with Nicholas of Cusa, 
who put much emphasis on quantification as a means to understand nature,
but concerning a theory for the basic structuring of the material world, he did also not advance beyond Plato.

While platonic thinking continued to be a pillar of western though in general, and western philosophy in particular,
and while many early natural philosophers and then scientists where deeply influenced by it through their scholastic education,
modern science has turned away from it step by step,
or maybe it was turned away by a platonic tradition refusing to pick up the questions which scientists had.
And although objective idealist thought went through several high-points afterwards,
the exchange with the rapidly developing natural sciences seem to have increasingly turned into two separate discussions within parallel universes.

Concerning for instance the Cambridge Platonists, 
More's {\it Hylarchic Principle} and Cudworths's {\it Plastic (i.e. forming) Life of Nature}
re-phrased the idea of a `world-soul', part of which is responsible for the `upkeep' of the material world,
but they did not come up with a detailed proposal for this process
(and maybe could not have done it at this time in the development of science),
which could have been taken up by the natural philosophers of their time.

And while German Idealism --  and especially (the traditional view of) Hegel's Philosophy -- can be understood
to be a refinement, perfection or even completion of the project
to find a most unified picture of the non-material layers of the traditional idealist construct,
no further reapproachment with science was achieved.
With Kant idealism took on its name and perfected it's role as a project to
explain (amongst others) the fundamental conditions, opportunities and likely or even necessary pitfalls of science,
in need of little interaction with actual science
(like Fichte's {\it Wissenschaftslehre} or Schelling's {\it Naturphilosophie}).
Concerned with only matter, science did not seem in the position to contribute much
to the philosophical fight about idealism vs materialism.
This of course back-fired with Moore's {\it Refutation of Idealism} and Russellian monism,
leading to a steep decline of interest in idealistic thought.

Although I am here concerned with objective idealism,
it needs it be mentioned that also subjective or critical accounts of idealism
such as those of Berkeley, Hume and Kant brought many ideas to objective idealism;
the most important for our discussion here is the rejection of
the substance theory of objecthood by Hume, who came up with the idea of bundle theories as alternative
(see below for a more detailed discussion).

Other philosophies with strong idealist leanings notably includes Leibniz',
who clearly tried to close the gap between idealism and science with his Monadology,
as well as the Leibniz-inspired thinkers of nineteenth century panpsychism,
part of which even went under the label `idealist panpsychism'.
From Leibniz we can extract the idea that a population of singular `souls' can
serve as an alternative to a single world-soul, 
but no detailed model for the emergence of the material world,
as `pre-established harmony' cuts all further questions short.
(Why are his monads causally impotent and therefore require a pre-established harmony?
Maybe -- long before evolutionary theory -- he couldn't imagine such a giant feat of self-organization,
or maybe it was beyond questioning that this was god's job?)

Idealist panpsychists have now and then taken up Leibniz' implied finding
that solving the mind/matter problem might simply require basic building blocks of non-material nature.
And while `atomistic' panpsychism (where non-material building blocks are assigned
to micro-scale physical entities) can not be seen as objective idealism,
some versions of panpsychism (where micro-scale physical entities are not seen as fundamental)
could be understood as such. (See below for a more detailed discussion.)
An especially interesting nineteenth century idealist panpsychist is Fechner,
not only because he actually came up with psycho-physical laws
(amongst others the one named after him, concerning the stimulus and intensity of a sensation,
giving a first, phenomenological, though not causal psycho-physical relationship),
but also because he genuinely tried to bridge the gap
when idealism and science broke over the theory of atoms,
by showing that atomism is not bound to materialism.
His {\it two aspect theory} nevertheless makes no advancement concerning the details of emanation
and his `universal law of psycho-physics' that everything mental is also physical
is a step back at least from the viewpoint of an idealist.
Later thinkers in between panpsychist and idealist positions
like James, Royce and Peirce retreated again to the discussion of general principles
rather than concrete interfaces with science.
Following the nineteenth century wave of panpsychism, the search for monistic philosophies often led to theories with idealist elements,
amongst others Russellian monism, variants of which have surfaced again more recently, e.g. with Goff.\cite{Goff2017}
An especially interesting case is North Whitehead, as his 
idea to turn from fundamental building blocks to fundamental processes
has clearly opened up a complete range of possible new theories.
Nevertheless, maybe due the rather abstract nature of his thinking,
to the best of my knowledge,
no scientific theory based on his ansatz has been created up to now.

Idealist and panpsychist thought clearly have a great overlap,
and accordingly, many idealistic ideas were put forward as part of panpsychist theories,
especially after Moore and Russell discarded many classical arguments for idealism.
Meixner distinguishes four versions of panpsychism based on their dualist or idealist and atomistic or holistic nature
(of which the dualist versions work no better than simple dualism for the mind/body problem,
and the atomistic versions suffer from the combination problem).\cite{Meixner2016}
Idealism is accordingly often treated as some form of idealist panpsychism,
but I think this is misguided:
Panpsychism is essentially based on the same notions of space and passive mental causality as materialism,
while idealism implies emergence of space and real agency (see below for my arguments).
Panpsychism is in this sense situated between materialism and dualism, i.e. still in the materialist's space and with subjects passively build up,
while idealism tries to take over using the hard shoulder, with dualism now a middle ground between materialism and itself.

Ongoing interest in panpsychism nevertheless proved valuable also for idealism,
in the sense that although classical idealism often continued in its science-alienated and science-alienating way,
work on idealist panpsychism also served objective idealist purposes.
I think it is fair to say that in recent decades work on panpsychism 
has shown that it has a fair chance to generate psycho-physical laws for how the mental is attached to the material but not vice versa
(take for instance the work of Lewtas\cite{Lewtas2017} 
which from the idealist view is an unattractive, one-sided solution to the interaction problem).

In more recent times, Sprigge and Foster have defended a panpsychist version of absolute idealism,\cite{Sprigge1984,Foster1982}
and with the return of metaphysics to philosophy departments and science's unaltered inability to explain consciousness,
more and more (mostly atomistic) panpsychist philosophies bring with them idealistic elements.\cite{BruntrupJaskolla2016,GoffMoran2021}
Concerning their chance for having impact on the scientific discussion,
panpsychists (especially atomistic dualist ones) have the advantage of being able to more easily build their theories on top of science,
without major re-interpretations of the later, while objective idealism -- in need of a much `richer' non-material world --
can hardly avoid such re-interpretations.
But although contemporary philosophers have taken up the challenge
to defend idealism against materialism again,
\cite{Idealism2017,HosleMuller}
even concerning idealism in the philosophy of science,
\cite{IdealSci2020}
no `mechanisms' of emanation have been proposed yet.

The above was of course only a very brief outline of some lines of idealistic thought,
but I think it is fair to say that although idealism has been a major stream of thought in philosophy since the very beginning,
no consistent formulation of idealism has emerged that allows for a direct engagement with modern science.
(To the best of my knowledge, a proper history of emanation is still missing,
though I think it would be a much-needed contribution to the idealistic venture.)
On the surface it looks as if idealism went out of fashion more or less in lockstep with the systematization of science according to our modern understanding of it,
as the views of the main protagonists became increasingly concerned with only their side of the medal,
and scientific successes allowed materialism to take over the public discourse.
Thus, while modern philosophy was deeply influenced by idealistic thought,
and panpsychism kept the connection and added new insights,
idealism itself retreated into the ivory tower:
Idealists pondered the fundamentals, with scientific findings as some kind of partial perspective of the fuller idealist world view,
but few theories were put forward how this could work out in detail and none at a level corresponding to modern science.
This of course is an exaggeration; it surely felt different to the idealists themselves.
Before the advent of modern science there was little need for a more detailed model of emanation,
which makes it even more amazing that Plato saw the need to have one.
Later on, scientific corner-stones like atomism were themselves very much open to debate
until the beginning of the twentieth century.
And the quasi-religious project of the idealistic unification of philosophy outlined by Plotinus and Proclus
which came to a halt with Hegel was anyhow an end in itself, quite independent of its (in-)ability
to connect to science and the material world in more than the most general way.
Likewise does modern physics in its quest for unification in the form of a {\it theory of everything} have very little worries about the mind/body problem.

Taking stock we find that after Plato the historical discussion continued at the level of very abstract organizational principles,
but without bridging concepts. Notable exceptions include (amongst others) Leibniz, Fechner and North-Whitehead,
of which unfortunately only Fechner's panpsychistic idea of psycho-physical laws is easily relatable to modern science.
More recently, and further building up on the works of  Leibniz and Fechner, panpsychists are coming closer to develop
a system of such laws, which can nevertheless not be considered as `rich' enough for a proper idealist system,
so in the end, the project of developing a scientifically tenable objective idealism is pretty much at the very beginning:
A lot can be taken from (neo-)platonic theories, for instance when they give us examples for how ideas could give form to matter (shapes to elements)
and how agency could be behind the upkeep of an idealist world.
Many ideas are available from Leibniz, for instance the idea that this upkeeping can be entrusted to a population of simpler minds instead of a single god-like mind.
And luckily, the contemporary discussion of panpsychism remains to be a rich source of new ideas for idealists,
as of course are recent works on (non-panpsychist) idealism
(some of which I will have the chance to discuss in the corresponding paragraphs below).
After this brief overview we will now turn to the `design questions' which we have to address to specify a theory of emanation.
\\ \\

\noindent {\bf \fontfamily{phv}\selectfont{What is the role of agents?}}\\
\noindent As we have seen above, objective idealism takes mind-independently existing non-material building blocks and at least on agent of some form as its fundamentals.
Different views of objective idealism can thus be differentiated first all by the role the agents play.
(Neo-)Platonistic thought considers active human agent and needs at least one god-like agent to take care of the world.
Leibniz considers agents across all scales, but their effectively passive nature puts his system in need of god as caretaker.
Hegel's absolute idealism dissolved the subject as agent into a sum of non-material elements in a larger World-Soul.
Holistic idealist panpsychism shows some similarity to this, but positions everything in the space of the material world,
while atomistic idealist panpsychism is letting go of the cosmic soul.
The core decisions we have to face are:\\
1. Who is responsible for the upkeep of the material world?
A population of singular agents -- parts of which could be of very simple nature, resembling cellular automatons -- or a god-like mind, or both?
(I will speak of `population', `god-like mind' or `mixed' theories in the following.)
Material causality is not an option, as in proper idealism the dispositional powers of matter are emanating from the non-material world;
material causality ceases to be a tool of explanation, but is itself in need of an explanation now.
In addition, we do not seem to observe a `mechanics' of thoughts, which could be emanating as material mechanics,
but the alternative of (add-on) psycho-physical laws is clearly an option,
only that this is rather a panpsychist than idealist project
(the results of which might nevertheless be importable to the later).

Unlike atomist panpsychist theories, population theories can prevent the combination problem
(of how human subjectivity arises from mere `mind dust'), if they allow agents to act on all scales,
(simple ones on the micro-scale, more complex ones at our meso-scale), because of emergent space, see below for a detailed discussion.
 Like holistic panpsychist theories, god-like mind theories run in the opposite of the combination problem,
 having to explain why we do experience ourselves as singular subjects and not part of a larger mind.
 Combined theories featuring a population of singular agents plus a god-like mind have no direct argumentative benefits
 over population-only theories, but are equally fine.
(Our experience as singular subjects is no argument against an additional larger mind,
in the same way as micro-scale agents would have no idea of our meso-scale existence.)
We should therefore start with a population theory, but are free to add a god-like mind afterwards.
Whether one wants to combine the two in some form thus remains to be first of all a question of faith,
i.e. as Alvin Plantinga put it, whether one takes religious belief as a proper basic belief.
\\
 2. The second question is closely related to the first question:
 If existent, are singular agents active or passive,
 i.e. are they just a bundle of non-material building blocks (e.g. sensations, thoughts, etc.)
 or do they command unique (though probably very limited) agency?
In general, active agents are preferable, as they can explain not only subjectivity, but also proper agency,
and they are available in population or combined theories.
The later ones allow also for passive agents,
which in turn allows for a holistic unification; simpler agents can then be parts of the cosmic mind.
There is a strong tradition in idealism for holistic unification,
but from the argumentative point of view it is more of a quasi-religious idea.
Quantum holism can probably be put forward as an argument here (though it is first of all a purely theoretical concept),
but biological evolution could equally well be seen as an argument for population theories. 
In summary I think there a no strong arguments for passive agents,
but a very strong one for active agents, namely that humans experience agency.
(While the hard problem of consciousness is nowadays accepted as a genuine problem,
there seems to me little that distinguishes the problem of agency from the one of qualia.)
A build-up of proper agency from passive agents at least does not seem to be a working option.\cite{Howell2015,Lewtas2017b,Lewtas2018}
\\ \\

\noindent {\bf \fontfamily{phv}\selectfont{Spacetime or space and time?}}\\
\noindent Assuming non-material objective existence beyond space, existence beyond time is an implied feature of idealism.
Time emanates then from the clockwork of the material world, where for proper material causation agents have to wait for the outcome of the actions of other agents.
(Due to their partly non-material nature, living beings could nevertheless have a very subjective experience of time, not always in line with physical time.)
I look at the question of whether this concept of time can give rise to a scientific theory like (general) relativity
in a forthcoming separate article on motion and relativity in objective idealism. 
For now I will continue supposing that this is indeed the case, but ignore relativity for now.
Modern science itself has two distinct ideas of time in general relativity and quantum theory,
with the later being based on the assumption of universal time, often with ad-hoc relativistic corrections.
Choosing between the two, the inability to do experiments on cosmic scales
gives some support to the idea to first start with a universal time conception of science first,
and assume that the effects of relativity are in turn impressed on space.
(Recent scientific attempts to unite general relativity and quantum theory often follow the same route.)
Apart from the above, the idealist can be agnostic of most other questions within the philosophy of time, I believe.
\\ \\

\noindent {\bf \fontfamily{phv}\selectfont{How does space emerge?}}\\
\noindent Unlike many current view points on panpsychism, which focus on the `narrow' mind-body problem of integrating consciousness into existing science,
idealism is usually held to account for the postulated objective existence of non-material entities like numbers, values, etc.,
i.e. the `broader' mind-matter problem of integrating the material with a proposed non-material world.
And unlike some proposed versions of panpsychism, a scientifically tenable idealism can therefore not be constructed
from adding non-material building blocks positioned in space(time) to the scientifically known constellations of material particles (particle-like field excitations),
as it would remain unclear how the objective existence of such entities as values or numbers would be situated in space.
(I have argued elsewhere in more detail that space has to be an emergent, not a basic feature of the idealist world.\cite{MKspace}) 
Science nevertheless seems to tell us, that the realization of spatial relations (unlike at least some non-spatial relations)
needs to emerge on the resulting micro-scale.
What is then needed seems to be some form of `static relationalism' which essentially works like substantial space.
A core question already for Leibniz and then Fechner, how space can arise from the relationship between spaceless entities,
needs to be answered for this. Again only agency comes to our help:
Space has to be the consequence of agents acting accordingly on certain properties of points in a network of relations.
(See below for a first suggestion on how this could work.)
\\ \\

\noindent {\bf \fontfamily{phv}\selectfont{How does material causality emerge?}}\\
\noindent As already outlined above, a second fundamental challenge to the needed re-interpretation of science concerns the role of causality and natural laws:
While at least some panpsychists can build their theories on top of the existing scientific structure of causality and natural laws,
the idealist is right from the beginning forced to render any `interaction problem' between mind and matter impossible:
Proposing the non-material as basic, makes matter, like space, an emergent feature of the world.
This leaves no room for a real problem of mind/matter interaction, but essentially precludes also material causality;
our laws of nature do not work on the non-material realm (thoughts don't hit on each other) and are thus also not basic,
so that now we need to explain how and why they emerge in the material world.
If the idealist wants to avoid the scenario of a pre-established harmony, i.e. choreographed change without real interaction in no need of further causal explanation,
mental causation comes into play as only known initiator of change in the non-material and -- because of its emergent nature -- thus also the material world.
As a result, idealism can make use of material causality and natural laws, but only as features emerging from a non-material world and agency.
This in turn requires at least one subject, with the ability to perceive and act on the non-material
(i.e. unlike in the discussion by Kant and German idealism, where the subject finally becomes superfluous as the nexus of perception,
here the subject is recovered as a necessary nexus of agency, which in turn requires perception.) 

Related to this is the problem that idealists, unlike modern science, can not resort to causal development based on randomness and natural laws
to explain the structure of the actual world, especially in the case of the big bang theory and biological evolution.
As material causation has to emerge from mental causation, only the evolution of either a population of subjects or a god-like mind
building up the emanating material world remains as `scientific' explanatory device,
quite in line with Goff's idea of a self-designed universe. \cite{Goff2019b}
\\ \\
\\

\noindent {\bf \fontfamily{phv}\selectfont{Which ontological model to take?}}\\
\noindent With all of the above in mind, and as we leave our common scientific picture of particles (or particle-like excitations of fields) in space (or spacetime) behind,
the question of course has to be, how we then manage to speak of things and change in the world.
(The choice of `things' as referentially basic could probably be justified with arguments from Strawson's Individuals.\cite{Strawson})
The philosophical tradition has developed essentially two routes to answer this question: Substance and bundle theory.
In substance theory, objects are discussed as to be constituted by a substance which bornes properties, while in bundle theory the object is just the bundle of its properties,
without a so-called `bare particular' at its core to identify its essence under change. While the idealist can in principle be agnostic of whether
objects are bundles of non-material building blocks of qualitative nature, or similar bundles with an additional bare particular at its core,
I have argued elsewhere that modern physics gives some unexpected support to the bundle theory view of objects.\cite{MKquantum}

A major issue with bundle theories is that the so-called `compresence' relation, which constitutes the bundling of qualities, leads to a range of logical puzzles.
The idealist is of course free to simply accept the imminent para-logical nature of the non-material world, as long as he can show that 
`material consistency', i.e. the sum of strict rules which are held up for the material world, prevent any spill-over of `weirdness' across the mind/matter gap.
It is nevertheless commonly assumed that this can still not work out, because in what is probably the most important argument against bundle theory,
it can be shown that compresence can not account for a proper identification of indiscernible objects in the material world:\cite{HS}
If position in space is not available as a feature, objects with exactly the same bundle of universal qualities become essentially the same object.
And if position in space should be invoked to account for this problem, it is unclear how the compresence relation can do this without infinite regress,
as same form of linkage seem to be required to make the relations of indiscernible objects consistent (see Ref. \cite{HS}
for further discussion).
While usually considered a knockout argument against bundle theories, I have argued elsewhere\cite{MKquantum}
that this bug is actually a core feature of a scientifically tenable objective idealism and that it even allows to shed new light on quantum theory.
Apart from the bundle theoretic one, other ontologies are in principle possible,
but none seems to be an equally good fit for the overall project, with the notable exception of North Whitehead's process philosophy maybe.
\\ \\

\noindent {\bf \fontfamily{phv}\selectfont{Bundle theories, mereology and space}}\\
\noindent An up to now (to the best of my knowledge) undiscussed problem with compresence nevertheless remains to be solved:
The idea of objects as bundles of qualities has a somewhat unclear notion of space and as a consequence also mereology.
Real objects are not simple, but spatially structured bundles of properties; a human person is a bundle of properties,
but also sub-bundles of properties, of which at least some, like for instance an arm, are spatially situated.
And to keep consistency with modern physics we have the additional requirement
that the consistent structuring of space must happen at the micro-scale,
as space seems to be fundamentally defined already at the level of elementary particles.
Such structuring is not properly accounted for in a simple notion of compresence,
which would have to additionally allow for the bundling of bundles and take into account some spatial positioning within bundles.
The idealist has the additional restriction (or if one thinks about it, actually opportunity) of emergent space,
so that compresence including sub-bundling must be understood as to be fundamentally constituted beyond (before) space.
Spatial situation (and thereby individuation) would then come into play from the inclusion of (sub-)bundles with spatial properties,
i.e. those bundles for which the stricter rules of the material world are obeyed.

As an example for this we can take a look at a person: She is a bundle of a mind and a body, partly material and partly non-material.
If we look at the `sub-bundle' which is only her arm I can identify further sub-bundles like the skin on her arm etc.
These bundles are largely material, but not completely so, in that they still have qualities like color etc.
If we continue to `un-bundle', we arrive at more and more materially composited building blocks,
but only at the very end we arrive at bundles with no non-material properties anymore, only defined by their material -- i.e. first of all spatial -- relations to each other.
(In analogy to Plato's fundamental bodies of geometric nature.)
But the fact that we call these relations material is only because they obey certain consistency requirements in spacetime;
if we could unbundle them further, they would just vanish into their non-material building blocks.
\\ \\
\\

\noindent {\bf \fontfamily{phv}\selectfont{The model: A bundle theoretic view of objective idealism}}\\
\noindent Based on the arguments above (on holism, space, matter, material causation, causal development and the problems of compresence),
the resulting model can be outlined like this:\\
\noindent 1. The world consists of non-material building blocks beyond space (which for our purpose here is not yet formed). These are of qualitative nature, which here means to include basic thoughts, mathematical entities, bare values etc. The exact nature of many of these classes of building blocks require further investigation, but does not need to be decided at this point to move on for now.\\
2. This world is inhabited by very simple, non-material subjects, able to perceive and act on the non-material building blocks. They are much simpler than what we commonly understand to be subjects, souls, etc. Let's call them agent modules.\\
3. Non-material building blocks can be (re-/de-)bundled by agent modules, thereby bringing diversity (including objects) into and change upon the world.\\
4. Between certain bundles very strict rules of manipulation are obeyed by the agent modules. The whole of these bundles represent the material world. The rules that are obeyed lead us to find laws of nature for the material world. Because space is emergent, agent modules can manipulate the world at very different spatial scales, but their ability to perceive and act depends on which material and non-material `machinery' is available to support them in these tasks: Very simple agents will consist of little but the core agent module and a few acquired non-material features and will account for the `causal upkeep' of the physical world at the micro-scale by acting out the rules underlying our natural laws. Such agents are no life forms as we know them, but should rather be seen as something like cellular automatons.
(In Physics, Wolfram has made the suggestion to explain our natural laws via the activity of cellular automatons.\cite{NKS})
More complex agents will grow into living beings of increasing complexity, depending on what non-material and material qualities are bundled with the non-material agent module at its core.
\\
5. The not only biological, but also physical world as we know it -- including space and matter, as well as the mind/matter gap -- developed as a product of the evolution of an agent module population. Worlds with unsustainable rules sort themselves out; if we would not have been lucky, we would not be here to wonder about it.
The material world nevertheless continues to function as the anchor of the non-material world,
with identity through positioning in space and material change through the motion of matter(-properties).
\\ \\

\noindent {\bf \fontfamily{phv}\selectfont{The model: The material world}}\\
To make this ansatz accessible to scientific investigation, we now have to specify in more detail how we have to understand space, material objects, properties and forces:\\
\noindent 1. Space is a network of bundles, each with the sole property of compresence of spatial relations to their neighbors in the network,
with these spatial relations being no more than some kind of `next' pointer.
Due to the indistinguishability of indiscernibles, this network of space `points' will encounter strong holistic effects,
but to investigate whether this can be a proper description of space(time) including (general) relativity is beyond the scope of this work.
I will try to address this in a forthcoming manuscript on motion, relativity and objective idealism,
but for now I will even make the additional assumption of time to work
in a naive way, as does quantum theory -- notwithstanding the fact, that this is incompatible with general relativity.\\
2. Objects are bundles of non-material building blocks of qualitative nature.
When we conjecture the existence of objects from rational inquiry, we might nevertheless identify pseudo-objects without objective existence with relevance for the material world. (Think of phlogiston, virtual particles, etc., which exist in the idealist world as non-material `idea' but not as functional objects in the material world).
A spatial bundle can be turned into an elementary particle if the compresence relation is extended beyond the spatial relations
to additional properties like mass, charge, spin, etc.\\
3. Discussing the scientifically relevant properties of objects we should assume these to be basic non-material building blocks.
As for objects, when we conjecture the existence of properties from rational inquiry, we might nevertheless identify pseudo-properties
without objective existence in the material world;
we especially have to discuss the role of speed, acceleration etc. Physical properties like mass or charge would then be non-material qualities bundled into the now material particle bundle.
\\
4. All forces must be considered pseudo-forces, as material causation emerges from agency.
The properties of mass or charge can therefore not be the underlying cause for gravitational or electromagnetic attraction or repulsion.
The realization of physical phenomena is due to the agency of micro-scale agents,
but these stick to their evolutionary acquired rules for how to act on their surrounding bundles,
thus letting us observe natural laws.
Particles with mass or opposite charge are forwarded to each other by micro-scale agents,
with mass or charge not so much the cause of their attraction,
but as the framework conditions of micro-scale agency;
properties don't cause, but allow for systematic causation.
Like with Lego blocks, structures appear according to certain rules,
but these rules are not the direct consequence of the brick's properties,
but of the actions of an agent acting upon them.
Especially in non-equilibrium situations, the situation at the micro-scale is wide open
for how exactly a specific physical process is realized, i.e. for fluctuations.
The forwarding (movement) of particles happens via the cleavage of existing relations and the formation of new ones,
so that in the end, all interactions arise from the forming or cleavage of (compresence) relations.
Interestingly, such a realization of physical forces shows parallels to process philosophy:
Material effects do not arise directly from properties (or substance), but from the process of agents acting upon them.
\\ \\

\noindent {\bf \fontfamily{phv}\selectfont{The mind/matter problem}}\\
\noindent This brings us back to the mind/matter problem:
At the meso-scale, a person can re-bundle some relations of the bundle that she is, i.e. she can act,
which will in turn influence also the material part of her that is her brain, from there cascading down via material causality
to some desired outcome in the material world.
It should nevertheless be clear that this can happen only insofar the framework conditions allow for this,
concerning the availability of appropriate mental facilities and possible physical states.
The brain -- probably via the default mode network -- would have to be seen
as first of all enabling different physical states with equal energetic and entropic properties
to allow for different material outcomes. I will try to further illuminate the implication of the outlined bundle-theoretic view of objective idealism
for neuroscience and the philosophy of information in a forthcoming manuscript.
\\ \\

\noindent {\bf \fontfamily{phv}\selectfont{Scientific questions to answer}}\\
\noindent With the above model in place, the work only starts. First of all the idealist now has to show that his theory
can indeed reproduce modern science not only on the conceptual level,
but down to it's mathematical machinery. For this the following questions need detailed answers:\\
1. Does the model allow for a mathematical model of time and space in line with (general) relativity?
(Here we have taken this as granted, but a proper demonstration is necessary.
As mentioned above, a first attempt will be made in a forthcoming manuscript on motion, relativity and objective idealism.)\\
2. Does the model allow for a mathematical consistent re-interpretation of quantum theory?
(First steps towards answering this question were made in a separate article.\cite{MKquantum})\\
3. Does the model allow for the import of our scientific concepts of energy, entropy, information, etc.?\\
4. Does the model allow for an integration with the standard models of particle physics and cosmology?
(Fine tuning and the somewhat arbitrary standard model parameters might be the results of an evolutionary formation of the cosmos,
but can this be brought in line with the mathematical machinery of modern physics?)\\
This research program is of course a moonshot project and might easily fail for many proposals,
but it is not completely different from established projects like the string theoretic re-interpretation
of quantum theory and general relativity. The most important point will therefore be to
make the model as precise as possible and thereby accessible to interested scientists,
with the goal of predicting and evaluating rationally accessible consequences of the new model.
Whether the final model will indeed allow us to address the open problems of physics, increase our understanding of the workings of the brain
or give us insight into the difference between human and machine intelligence remains to be seen.
\\ \\

\noindent {\bf \fontfamily{phv}\selectfont{Conclusions}}\\
\noindent
To stay relevant and to bring its potential for further growth to the table,
idealism will have to be more than one of many possible background stories for science.
For this, the idealist has to give a mathematically consistent re-interpretation of the physical world as a limiting case of a both material and non-material world.
In this article I have made a first attempt at such a re-interpretation, with a model based on a bundle-theoretic view of objective idealism.
Any theoretical construct which hopes to shed light on the mind/matter gap will ultimately be judged by what explanatory power it can provide and how useful it therefore is to explore reality. For this, no single argument will do, but only overall better fit.
The hypothesis here is that idealism has the potential to not only close the gap, but contribute significantly to our understanding of modern science.
I have made a first attempt at this with a new interpretation of quantum theory based on the above model elsewhere.\cite{MKquantum}
With proposals like this, idealism of course makes itself vulnerable - first of all against scientific rejection of certain theory parts.
But as with personal relationships, without vulnerability there can be no deeper connection.
In this sense, traditional idealism was probably too invulnerable to stay as relevant as it once was.
\\ \\





}
\end{document}